\documentclass{aa}

\usepackage{float} 

\usepackage{dcolumn}    
\usepackage{bm}         

\usepackage{amsmath} 
\usepackage{amssymb} 
\usepackage{amsfonts} 
\usepackage{latexsym}
\usepackage{enumitem} 

\usepackage{booktabs} 

\usepackage{natbib}

\usepackage[normalem]{ulem}

\usepackage[english]{babel}
\usepackage[expansion=all,babel]{microtype}

\usepackage[usenames]{xcolor}

\usepackage{mathtools}

\usepackage{multirow}

\usepackage{newtxtext,newtxmath}

\usepackage{natbib,twoopt}
\usepackage[hyphenbreaks]{breakurl}
\usepackage[breaklinks,hidelinks,colorlinks,citecolor=blue,breaklinks=true]{hyperref}
\bibpunct{(}{)}{;}{a}{}{,}             
\definecolor{cobalt}{rgb}{0.06, 0.2, 0.65}
\hypersetup{
	colorlinks,
	citecolor=cobalt,
	linkcolor=[rgb]{0.8, 0.2, 1.0},
	urlcolor=cobalt,
}
\makeatletter
\newcommandtwoopt{\citeads}[3][][]{\href{http://adsabs.harvard.edu/abs/#3}%
	{\def\hyper@linkstart##1##2{}%
		\let\hyper@linkend\@empty\citealp[#1][#2]{#3}}}
\newcommandtwoopt{\citepads}[3][][]{\href{http://adsabs.harvard.edu/abs/#3}%
	{\def\hyper@linkstart##1##2{}%
		\let\hyper@linkend\@empty\citep[#1][#2]{#3}}}
\newcommandtwoopt{\citetads}[3][][]{\href{http://adsabs.harvard.edu/abs/#3}%
	{\def\hyper@linkstart##1##2{}%
		\let\hyper@linkend\@empty\citet[#1][#2]{#3}}}
\newcommandtwoopt{\citeyearads}[3][][]%
{\href{http://adsabs.harvard.edu/abs/#3}
	{\def\hyper@linkstart##1##2{}%
		\let\hyper@linkend\@empty\citeyear[#1][#2]{#3}}}
\makeatother

\makeatletter
\def\@LN#1#2{} 
\def\@LN@col#1{} 
\let\linenumbers\relax
\let\nolinenumbers\relax
\makeatother

\usepackage{cleveref}

\definecolor{SC}{RGB}{199,21,133} 
\definecolor{MB}{RGB}{255,99,71} 

\newcommand{\sat}[1]{{#1}_{\mathrm{sat}}}

\begin{document}

\title{Thermal evolution of neo-neutron stars.} 
\subtitle{II. Temperature-dependent crusts}

\author{Sagnik Chatterjee\inst{1}\email{sagnik.chatterjee@nipne.ro}
\and Mikhail V. Beznogov\inst{1}\corrauth{mikhail.beznogov@nipne.ro}
\and Dany Page\inst{2}\email{page@astro.unam.mx}
\and Adriana R. Raduta\inst{1}\email{araduta@nipne.ro}}

\institute{National Institute for Physics and Nuclear Engineering (IFIN-HH), RO-077125 Bucharest, Romania
\and Instituto de Astronomía, Universidad Nacional Autónoma de México, Ciudad de México, CDMX 04510, Mexico}

\abstract{
The neo-neutron star phase is an intermediate stage in the evolution from a hot, lepton-rich compact object, which is formed in the aftermath of a core-collapse supernova or an accretion-induced collapse of a white dwarf, to a cold deleptonized neutron star (NS). Alternatively, this phase can occur after a binary NS merger if the final compact object does not collapse immediately into a black hole. Radial temperature profiles provided by numerical simulations of proto-NS evolution suggest that the composition and mechanical structure of the star's core at the beginning of the neo-NS phase are approximately identical to those of a cold NS. In contrast, the composition and mechanical structure of the outer layers (i.e., the crust) are expected to change as the star cools. The thermal evolution of neo-NSs with temperature-dependent (inner) envelopes was considered by \cite{Beznogov_ApJ_2020}.}
{Here, we investigate what happens if the equation of state (EoS) of the entire crust, including its composition, depends on temperature.}
{We analyze the thermal, mechanical, and chemical evolution of neo-NSs by further developing and employing \texttt{neo-NSCool}, our NS thermal evolution code.} 
{First, we show that thermally dripped neutrons slightly slow down the cooling during the early thermal relaxation stage. Then, we demonstrate that the EoSs featuring exotic light species in the deepest layers of the crust result in significantly slower cooling in the photon-cooling era compared to the EoSs that favor massive nuclei. The crust's composition also impacts the crystallization process as well as the contraction of the crust during cooling. The EoSs with light nuclei cause the innermost layers to remain liquid for a longer time than the EoSs with heavy nuclei.}{}


\titlerunning{Thermal evolution of neo-neutron stars}
\authorrunning{Chatterjee et al.}

\maketitle

\section{Introduction} 
\label{sec:Intro}

Neutron stars (NSs) are among the densest objects in the Universe. 
They represent the final stage of the evolution of ordinary stars following core-collapse supernovae (CCSN; \citealt{Baade_1934}), accretion-induced collapses of white dwarfs (WDs; \citealt{Canal_AA_1976}), or binary NS mergers~\citep{Kluznak_ApJL_1998}. 
In either of these scenarios, the precursor of what will become an NS is a hot, potentially lepton-rich and fast-rotating object, with convoluted radial profiles of density, temperature, entropy, lepton fraction, etc., all of which are strongly dependent on the characteristics of the progenitor star, microphysics and, in the case of mergers, details of the orbital dynamics in the inspiral phase. 

The evolution from a proto-neutron star (PNS), which is a central compact object remaining after the core of a collapsing massive star reaches suprasaturation densities and bounces, to a cold deleptonized NS is relatively well understood~\citep{Burrows_1986,Prakash_PhysRep_1997,Pons_ApJ_1999,Pascal_MNRAS_2022}. 
It basically consists of three stages.
The first stage corresponds to the fast contraction of the PNS from a radius of about 150~km to about 20~km, due to neutrino leakage from the low-density outer layers, which are thus deprived of pressure support.
The next stage is characterized by neutrino diffusion from the central region of the star to the outer shells, which results in core heating and deleptonization.
The third stage corresponds to the overall cooling of the PNS; during this epoch, both the entropy and its gradient steadily decrease.
The latter two stages are estimated to last for a few tens of seconds~\citep{Pascal_MNRAS_2022} or for several tens of seconds~\citep{Pons_ApJ_1999}, depending on whether convection, which allows for efficient heat and neutrino transport from the inner regions to the neutrinosphere~\citep{Pascal_MNRAS_2022}, is accounted for.
The numerical simulations of PNS evolution in \citet{Pons_ApJ_1999} end when the PNS has a radial profile of temperature that decreases from $\sim 5$~MeV in the core to $\sim 1$~MeV in the outer shells and is transparent to neutrinos.
Had these simulations continued, the PNS would have cooled further, shrunk, and approached the cold catalyzed NS phase.
Models predict that the surface of the newborn NS will consist of iron-peak nuclei, but subsequent accretion of matter that failed to be ejected by the supernova explosion may deposit a thick layer of light elements on it \citep{2020ApJ...898..125P,2026MNRAS.548ag329B}.

The thermal properties of middle-aged ($10^2 - 10^6$~yr) isolated NSs are also well known.
This is due to both the availability of thermal surface emission data for two dozen stars and an array of dedicated numerical simulations~\citep{Schaab_ApJ_1998,Yakovlev_Uspekhi_1999,Gnedin_MNRAS_2001,Page_ApJ_2004,Yakovlev_RevAA_2004,Page_ApJ_2009,Tsuruta_ApJ_2009,Beznogov_MNRAS_2015a,Beznogov_MNRAS_2015b}.
For a compilation of observational data, see \citet{Potekhin_2020}.
It has been established that, once the star is thermally relaxed and for $\sim 10^5$~yr, the surface temperature is mainly determined by the neutrino emission from the core, which is intimately linked to the composition of dense matter and the superfluidity properties of various fermionic species. 
As such, studies of NS thermal evolution have the potential to reveal dense matter information complementary to that extracted from measurements of maximum NS masses~\citep{Antoniadis_Science_2013, Cromartie_Nature_2020}, radii of NSs with masses $1.04 \lesssim M/M_{\sun} \lesssim 2.08$~\citep{Salmi_ApJ_2022,Salmi_ApJ_2024b,Vinciguerra_ApJ_2024,Miller_ApJ_2026}, and tidal deformability from the GW170817 event~\citep{Abbott_PRL_2017}, all of which probe only the mechanical structure, but not the composition.

\looseness =-1
Numerical simulations of long-term thermal evolution, including those cited above, assume that the mechanical structure and composition of NSs correspond to those at zero temperature.
Given that for $t \gtrsim 0.3$~yr the local temperature does not exceed $\approx 0.2$~MeV, this approximation is highly accurate.
When addressing earlier stages, where the NS temperature is of the order of $1-2$~MeV, the fixed structure and fixed composition approximations become questionable.
First, finite temperatures increase the pressure in the low-density part of the equation of state (EoS), which translates into EoS stiffening and results in crusts that are thicker than those at $T=0$.
Furthermore, the high temperature sensitivity of the electron chemical potential means that the beta-equilibrium at finite temperature is achieved for values of the proton fraction other than those at zero temperature.  
It is clear that the neo-NS phase, which begins when the PNS becomes transparent to neutrinos, i.e., after $\approx 30-60$~s after the core bounce, requires improved treatment.

A first step in this direction was presented in \citet{Beznogov_ApJ_2020}, hereafter dubbed Paper~I. 
In that paper, the authors allowed the (inner) envelope, which at $t=0$ corresponds to the density domain $10^5 \leq \rho \leq 10^{11}$~g/cm$^3$, to shrink while the NS cools down.
They examined how long the star can sustain Eddington luminosity, how its inner envelope evolves with time, and, among other things, how the initial surface luminosity affects the star's contraction.

The aim of this paper is to further develop the treatment of neo-NSs that contract as they cool down.
In addition to what was done in Paper~I, we now also allow for temperature effects in the entire crust up to the crust-core transition.
Here, we consider only the case of nucleonic NSs with $M=1.4~M_{\odot}$.
The remainder of the paper has the following structure. 
In Sec.~\ref{sec:formalism}, we review the formalism.
The initial luminosity profiles and the  computer code used for numerical simulations are presented in Sec.~\ref{sec:setup}. 
Section~\ref{sec:EoS} is devoted to crust models. 
Results on the thermal evolution, contraction of cooling NSs and crystallization of the crust are offered in Sec.~\ref{sec:result}.
In Sec.~\ref{sec:Concl}, we draw the conclusions.

\section{Formalism}
\label{sec:formalism}

In this paper, we assume that the NSs are spherically symmetric and ignore the effects of rotation, magnetic fields, and accretion.
For thermal evolution simulations, we employ the formalism proposed in \citep{Potekhin_2018} and in Paper~I.
In the following, we provide a brief description and refer interested readers to the original publications.

Thermal evolution is described by the following two equations:
\begin{align} 
  &\widetilde{L} = - K (4\pi r^2)n e^{\phi} \frac{\partial \widetilde{T}}{\partial a},
  \label{eq:Ltilde} \\
  &e^{\phi} \frac{\partial (\widetilde{T} e^{- \phi})}{\partial t} = - \frac{1}{C_\mathrm{V}} (\widetilde{Q}_\mathrm{L} + \widetilde{Q}_{\nu}+ \widetilde{Q}_\mathrm{V}),
  \label{eq:Ttilde}
\end{align}
where $\widetilde{L} = L e^{2\phi}$ and $\widetilde{T} = T e^{\phi}$ are the redshifted luminosity and temperature; $e^{2\phi}$ is the time component of the metric;
$n$ is the baryon number density;
$C_\mathrm{V}$ and $K$ stand for specific heat (per unit volume) and thermal conductivity, respectively; 
$r$ and $a$ represent the circumferential radius of the star and the enclosed baryonic number; 
$\widetilde{Q}_\mathrm{L}$, $\widetilde{Q}_{\nu}$, and $\widetilde{Q}_\mathrm{V}$ denote the heat (per unit volume) lost (or gained)  due to luminosity gradients, neutrino energy loss, and contraction energy, respectively; $t$ is the time. 
Since in our simulations the star structure evolves in time, we employ the enclosed baryonic number as an independent Lagrange radial variable.

At the center of the star, the luminosity, radius, enclosed baryonic number, and enclosed gravitational mass all vanish by definition.
The equilibrium configuration of the NS is determined by the value of pressure at the center.
The surface of the star corresponds to the photosphere \citep{Hansen2004StellarInteriors}.
Surface luminosity ($L_\mathrm{s}$) and temperature ($T_\mathrm{s}$) are related by
\begin{equation}
  L_\mathrm{s} = 4 \pi \sigma_\mathrm{SB}R^{2}T_\mathrm{s}^4,
  \label{Ls_eq}
\end{equation}
where $R$ stands for the radius of the star and $\sigma_\mathrm{SB}$ is the Stefan-Boltzmann constant.

The pressure on the surface ($P_\mathrm{s}$) is related to $L_\mathrm{s}$ by means of the Eddington (photospheric) condition \citep{Hansen2004StellarInteriors}:
\begin{equation} 
  P_\mathrm{s} = \frac{2}{3}\frac{g_\mathrm{s}}{\kappa_\mathrm{s}}\left(1+\frac{L_\mathrm{s}}{L_\mathrm{Edd}}\right), \quad L_\mathrm{Edd} = \frac{4\pi G M e^{\lambda}}{\kappa_\mathrm{s}},
  \label{Ps_eq}  
\end{equation}
where $g_\mathrm{s} = e^{\lambda} GM/R^2$ is the surface free-fall acceleration ($G$ is the gravitational constant, $M$ is the total gravitational mass of the star, and $e^{2 \lambda}$ is the radial component of the metric), $\kappa_s$ represents the Rosseland mean opacity on the surface, and $L_\mathrm{Edd}$ is the surface Eddington luminosity.
On the surface of the star, $e^{\phi (R)} = e^{-\lambda (R)} = \sqrt{1-2GM/Rc^2}$, where $c$ is the speed of light in vacuum.

The outermost layers of the NS constitute its envelope, which extends from the star's surface to a bottom layer characterized by some pressure $P_\mathrm{b}$ and a corresponding density $\rho_\mathrm{b}$. 
Since we are investigating isolated non-accreting NSs, and for consistency with Paper~I, we assume that the envelope consists of pure iron. Similarly, we fix the value of $\rho_\mathrm{b}$ at $t=0$ to $10^5$~g/cm$^3$.
However, as the star cools and shrinks, $\rho_\mathrm{b}$ will increase.
The thickness of the envelope is of the order of a few hundred meters or less.
The thermal relaxation timescale of this layer is much shorter than the cooling timescale of the interior, which allows us to treat it separately and in a time-independent way; see Paper~I for more details. 
The temperatures at the bottom of the envelope and on its surface are related by the ``$T_\mathrm{s}$-$T_\mathrm{b}$'' relation~\citep{Gudmundsson, Beznogov_PhysRep_2021}: 
\begin{equation}
  T_\mathrm{s}(T_\mathrm{b},g_{s,14}) = g_{s,14}^{1/4}T_\mathrm{s} (T_\mathrm{b},g_{s,14} = 1),
  \label{tstb_eq}
\end{equation}
with $g_{\mathrm{s},14} = g_\mathrm{s} / (10^{14}~\mathrm{cm~s^{-2}})$.
As in Paper~I, we assume that there are no heat sources or sinks in the envelope.
This is straightforward: there is no accretion; the envelope is made of iron, which implies an absence of thermonuclear reactions; and neutrino energy losses at densities lower than $10^5$~g/cm$^3$ and temperatures lower than $0.1$~MeV are negligible.
Consequently, the energy flux through the envelope is constant (i.e., $L_\mathrm{s} = L_\mathrm{b}$).

Hereafter, we adopt the common nomenclature, where the outflowing luminosity is written in terms of an effective temperature $T_\mathrm{eff}$ as
$L = 4\pi \sigma_\mathrm{{SB}} R^2 T^4_\mathrm{eff}$, with redshifted quantities defined as $L^{\infty} = e^{2 \phi} L = \widetilde{L}_\mathrm{s}$, $T^{\infty}_\mathrm{eff} = e^{\phi}T_\mathrm{eff}$, and $R_{\infty} = e^{-\phi} R$; for our study we have $T_\mathrm{eff} = T_\mathrm{s}$.

Finally, a note on the naming conventions is necessary to avoid misunderstanding. 
In Paper~I, the region with density $\rho < 10^5$~g/cm$^{3}$ was referred to as the ``outer envelope'', the region  $10^5 < \rho~[\mathrm{g/cm^{3}}] < 10^{11}$ was referred to as the ``inner envelope'', and the regions with $\rho > 10^{11}$~g/cm$^{3}$ were referred to as the ``crust'', ``inner crust'' and ``core'', depending on the density. 
Here, we will refer to the region with $\rho < 10^5$~g/cm$^{3}$ as the ``(outer) envelope'', the region with $10^5~\mathrm{g/cm^{3}} < \rho < \rho_{\mathrm{n;\,drip}}$ as the ``outer crust'', and the region with $\rho_{\mathrm{n;\,drip}} < \rho < \rho_\mathrm{cc}$ as the ``inner crust'', where $\rho_{\mathrm{n;\,drip}}$ and $\rho_\mathrm{cc}$ represent the neutron drip and crust-core transition densities, respectively;
their typical values are $\rho_{\mathrm{n;\,drip}} \approx 4 \times 10^{11}$~g/cm$^3$ and $\rho_\mathrm{cc} \approx 1.7 \times 10^{14}$~g/cm$^3$.
From a calculation point of view, the boundary between the envelope and the outer crust is the surface of the star.

\section{Setup}
\label{sec:setup}

In this paper, we focus on the thermal evolution of isolated, non-rotating NSs with $M_G = 1.4~M_\odot$, whose hadronic sector consists of neutrons and protons.
We prefer the canonical mass for several reasons.
First, this allows us to neglect the direct Urca processes, which are the most efficient cooling mechanisms and whose threshold densities ($n_{\mathrm{DU}}$) are highly uncertain.
Indeed, state-of-the-art investigations of the EoS of dense matter fail to constrain $n_{\mathrm{DU}}$ despite employing phenomenological nuclear energy density functionals, Bayesian statistics and a wide array of nuclear and astrophysical constraints~\citep{Beznogov_PRC_2023,Beznogov_ApJ_2024,Beznogov_PRC_2024,Cartaxo_ApJSS_2026}.
This situation stems from the scarcity of experimental nuclear information on neutron-rich dense matter, the poorly understood sensitivities of astrophysical data to the behavior of dense isospin-asymmetric matter, and the unknown composition of dense matter with densities of the order of a few $\sat{n}$, the latter being linked to the high-density behavior of the nuclear symmetry energy
($\sat{n} \approx 0.16~\mathrm{fm}^{-3} \approx 2 \times 10^{14}~\mathrm{g/cm^3}$ being the nuclear saturation density).
Statistical analyses of the thermal evolution of isolated and especially accreting NSs support the hypothesis that the direct Urca processes should operate in sufficiently massive NSs (with masses exceeding the canonical mass; see \citealt{Beznogov_MNRAS_2015b}).
Second, a moderate compactness prevents excessive compression of the crust and, thus, maximizes temperature-induced variations.
Finally, it allows a straightforward comparison with the results of Paper~I, which used the same $M_G = 1.4~M_{\odot}$.
In the following, we discuss the initial profiles of luminosity and temperature and the basic features of the computer code we employ.

\subsection{Initial profiles of local luminosity and temperature}
\label{ssec:LProfiles}

\begin{figure}
  \centering
  \includegraphics[width=\linewidth]{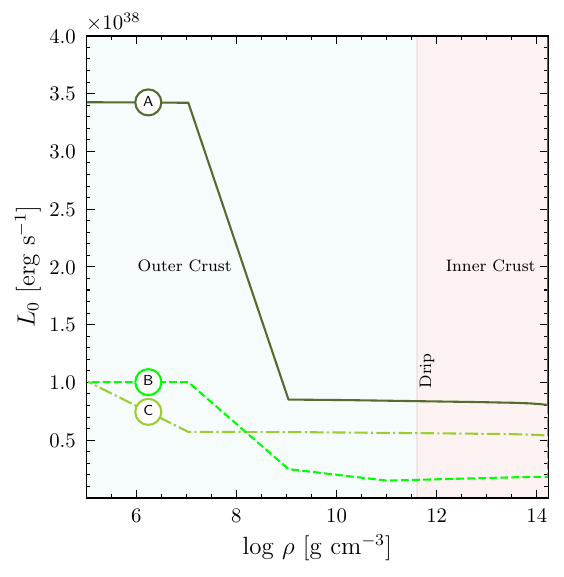}
  \includegraphics[width=\linewidth]{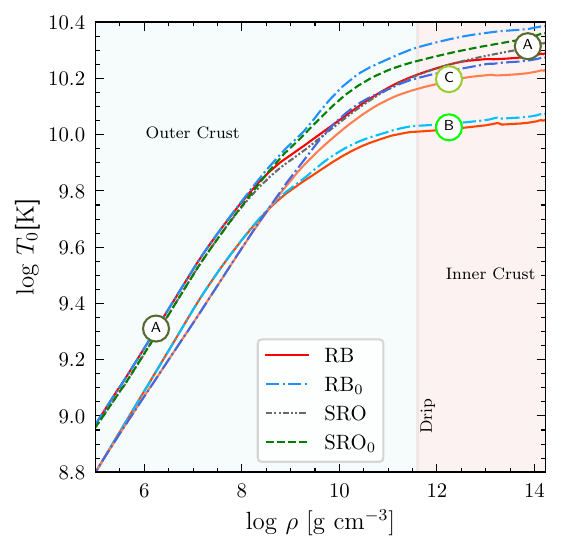}
  \caption{Radial profiles of the initial ($t=0$) local luminosity (top panel) and local temperature (bottom panel) as functions of density.
  Outer and inner crusts are indicated by background colors. The drip density, marked by a vertical line, corresponds to the cold HZD EoS~\citep{HZD}.
 }
  \label{fig:IniProfs}
\end{figure}

\looseness=-1
Since the neo-NS phase is the intermediate phase in the evolution from a hot, lepton-rich PNS to a cold, deleptonized NS, it is natural to assume that the local temperature profiles at the beginning of the neo-NS phase coincide with the local temperature profiles at the end of the PNS phase. 
Numerical simulations of PNS evolution~\citep{Pons_ApJ_1999,Pascal_MNRAS_2022} show that these profiles depend on the progenitor mass and a series of microphysics ingredients (e.g., neutrino interactions and EoS), as well as simulation technicalities (e.g., neutrino transport).
Moreover, PNS calculations usually do not extend far into the outer layers of the star; frequently, the data is plotted as a function of the enclosed baryonic mass.
Consequently, the thermal profiles of the crust cannot be extracted, as the crust contains a negligible amount of mass compared to the core.

Thus, in the absence of necessary ``starting'' data, we model the initial configuration of the neo-NS by means of arbitrary schematic luminosity profiles A, B, and C, depicted in Fig.~\ref{fig:IniProfs} (top panel).
In all cases, the maximum local luminosity corresponds to the bottom of the envelope and is the same as the surface luminosity.
For case A, the local luminosity at the surface is $\approx 75$\% of the Eddington luminosity.
Ideally, we would have started with the Eddington luminosity at the surface, as in Paper~I; however, our code did not converge for surface luminosities exceeding 75\% of the Eddington limit.
For the same reason, profile A has a sharp decrease in luminosity starting at $10^7$ g/cm$^3$.
In two other cases (B and C), the local luminosity at the surface is much lower.

Complementary information is provided in the bottom panel in terms of the initial local temperature profiles that correspond to the initial luminosity profiles A, B, and C and to various considered EoSs of the crust; for the latter, see Sec.~\ref{sec:EoS}.
The $T_0$-profiles are obtained by solving an equation similar to Eq.~\eqref{eq:Ltilde}, which depends on the thermal conductivity and mechanical structure.
SRO$_0$ leads to slightly higher temperatures in the inner crust compared to SRO, suggesting that the former model has lower thermal conductivity than the latter.
The difference between RB$_0$ and RB is more pronounced, indicating more substantial differences in their thermal conductivities.
As we shall detail in Sec.~\ref{sec:EoS}, SRO$_0$ and SRO (RB$_0$ and RB) use the same $P(\rho,T)$ relation but differ in the composition of the crust.
For all cases, the deep layers are warmer than the shallow ones, and higher luminosities get translated into higher temperatures.
Also, had the high-luminosity plateaus of cases A and B been more extended, higher temperatures would have been obtained in the inner crust.

We fixed the initial core temperature to the value obtained at the crust-core boundary.

\subsection{Extension of NSCool}

We simulate the thermal evolution of NSs using an extension of the NS cooling code {\tt NSCool}~\citep{Page_NSCool_1d,Beznogov_ApJ_2020}.
In addition to what was done in Paper~I, here we evolve the entire crust over time, taking into account the temperature dependence of its mechanical structure and composition. 
Similar to Paper~I, this is done by solving the Tolman-Oppenheimer-Volkoff equations in the crust at each timestep in addition to the thermal evolution equations \eqref{eq:Ltilde} and \eqref{eq:Ttilde}.
As in Paper~I, only the thermal evolution equations are solved in the  core.

In both the core and the crust, we allow for all microphysics processes implemented in the standard version of {\tt NSCool} \citep{Page_ApJ_2009} except for the heat capacity of the crust, which is discussed in Sec.~\ref{ssec:Cv}.
For the (outer) envelope, we employ the model and ``$T_\mathrm{s}$-$T_\mathrm{b}$'' relations developed in Paper~I specifically for high-luminosity envelopes.
We disregard neutron and proton pairing throughout the star. 

\section{Equation of State Models}
\label{sec:EoS}

\renewcommand{\arraystretch}{1.1}
\setlength{\tabcolsep}{7.0pt}
\begin{table}
  \caption{EoS models for crust composition in $\beta$-equilibrium.}
    \begin{tabular}{l c l}
        \toprule
        Model                 & Effective interaction   &  Composition\\
        \midrule
        ${\mathrm{RB}_0}^{(a)}$   &  BBSk1$^{(a)}$  &  $T=0.1$~MeV \\
        ${\mathrm{RB}}^{~(a)}$  &  BBSk1$^{(a)}$  &  $T(\rho)$ \\
        ${\mathrm{SRO}_0}^{(b)}$  &  APR$^{(c)}$    &  $T=0.1$~MeV \\
        ${\mathrm{SRO}}^{~(b)}$ &  APR$^{(c)}$    &  $T(\rho)$ \\
        \bottomrule
    \end{tabular}
    \tablefoot{Model names are listed in col.~1, underlying effective interactions in col.~2, and the temperatures at which the compositions were computed in col.~3.
    \tablefoottext{a}{\cite{Raduta_AA_2025};}
    \tablefoottext{b}{\cite{Schneider_PRC_2019};}
    \tablefoottext{c}{\cite{APR}.}}
    \label{tab:Models}
\end{table}
\renewcommand{\arraystretch}{1.0}

In the following, we use the term ``EoS'' in a generic sense, referring to both the $P(\rho,T)$ dependence and the composition of matter at density $\rho$ and temperature $T$.

In all cases, we use the zero-temperature APR EoS \citep{APR} for the NS core. 
Employing a cold EoS for the core is not an issue, as the thermal energy in the core at the considered temperatures of 1-2~MeV is negligible compared to the Fermi energy.
For the NS crust, we alternatively use the EoSs by \cite{Schneider_PRC_2019} (hereafter dubbed SRO) and \cite{Raduta_AA_2025} (hereafter dubbed RB).
For $T \geq T_\mathrm{lim}$, the EoS information is extracted from the general purpose tables publicly available on the \texttt{CompOSE} online repository\,\footnote{\url{https://compose.obspm.fr/}}\citep{Typel_EPJA_2022},
where $T_\mathrm{lim}$ denotes the minimum temperature for which the data exist in the \texttt{CompOSE} tables.
For both SRO and RB,  $T_\mathrm{lim}=0.1$~MeV. 
For $T < T_\mathrm{lim}$ and $\rho < 10^{11}$~g/cm$^3$, we employ the $P(\rho,T)$ dependence from \citet{Potekhin_2010} and the composition from the corresponding general purpose EoS taken at $T_\mathrm{lim}$.
For $T < T_\mathrm{lim}$ and $\rho \geq 10^{11}$~g/cm$^3$, we use values from the corresponding general purpose EoS taken at $T_\mathrm{lim}$ both for the composition and for the $P(\rho,T)$ dependence, which, obviously, becomes $P(\rho,T =  T_\mathrm{lim})$ dependence.
In other words, we consider that for $T < T_\mathrm{lim}$ and $\rho \geq 10^{11}$~g/cm$^3$ the temperature dependence of an EoS can be completely ignored.

For densities lower than the crust-core transition density and temperatures lower than a few MeV, SRO and RB provide very similar $P(\rho,T)$-relations in spite of differences in modeling and the underlying effective interaction~\citep{Burrows_ApJ_1984}.
In contrast, the crust composition is highly model-dependent. 
In the first place, these differences arise from the pool of nuclei allowed by each model.
As such, \cite{Schneider_PRC_2019}, who rely on the liquid drop approximation for the nuclear mass, predict that, at densities slightly lower than the crust-core transition density, the crust is made of extremely massive and neutron-rich nuclei, while \cite{Raduta_AA_2025}, who use experimental and/or evaluated nuclear masses, favor less massive nuclei with limited isospin asymmetry.
Differences in composition also arise from the way in which nucleonic pairing and shell effects are implemented in the nuclear masses, but these effects are subdominant and rapidly washed out at finite temperature.

In the crust, finite temperature effects modify both the $P(\rho)$-relations and the composition. 
For a given temperature, thermal effects increase as the matter becomes more dilute and are more significant for pressure than for energy density~\citep{Raduta_EPJA_2021}. 
In addition, the strong temperature dependence of the electron chemical potential entails a non-negligible modification of the proton fraction of the $\beta$-equilibrated matter.
The most notable model dependence of crust composition at finite temperature manifests whenever exotic light species are included in the nuclear pool. 
An example in this sense is provided in \citet{Beznogov_PRC_2026}, which used the RB EoS, and where the deep layers of the inner crust appear to be made of $^{14}$He.
Though it remains to demonstrate to what extent such species survive in-medium effects, it is easy to understand that they will modify the thermal properties of the crust and its crystallization. 
The latter aspect will be addressed in Sec.~\ref{sec:Crystal}.

The recipes used to portray the composition of the crust are listed in Table~\ref{tab:Models}.
The comparison between the cooling simulations that account for the temperature effects on the crust composition (models RB and SRO) is meant to highlight the composition dependence, including the role of exotic light species.
Further comparison with the results of simulations with fixed crust composition (models RB$_0$ and SRO$_0$) aims to gauge the importance of the ``real-time'' modifications of the composition.

\subsection{Crust composition}
\label{ssec:CrustCompo}

\begin{figure}
    \centering
    \includegraphics[width=1.0\linewidth]{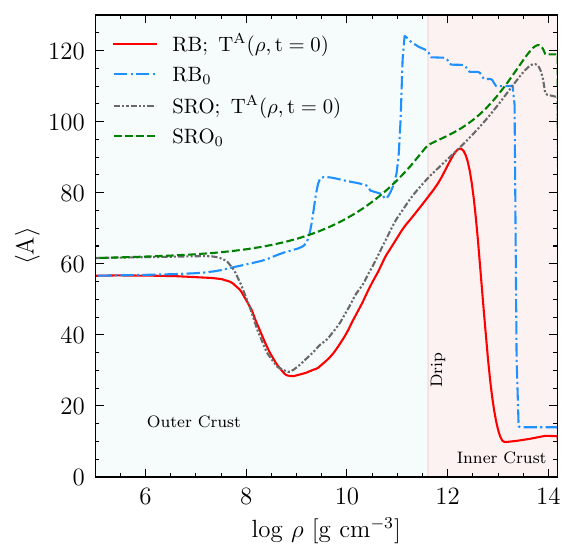}
    \includegraphics[width=1.0\linewidth]{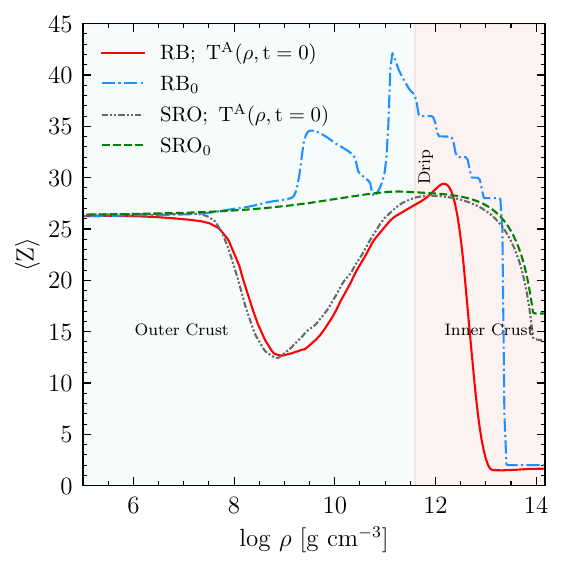}
    \caption{Composition of the neo-NS crust in terms of the average mass number (top panel) and charge number (bottom panel) as a function of density at $t = 0$.
    For RB and SRO, we employ the $t=0$ temperature profile A from Fig.~\ref{fig:IniProfs}, while for RB$_0$ and SRO$_0$, the composition is computed at $T=0.1$~MeV.      
    For the theoretical approaches and local temperature profiles, see Table~\ref{tab:Models}.}
    \label{fig:Compo_avAZ}
\end{figure}

\begin{figure}
    \centering
    \includegraphics[width=1.0\linewidth]{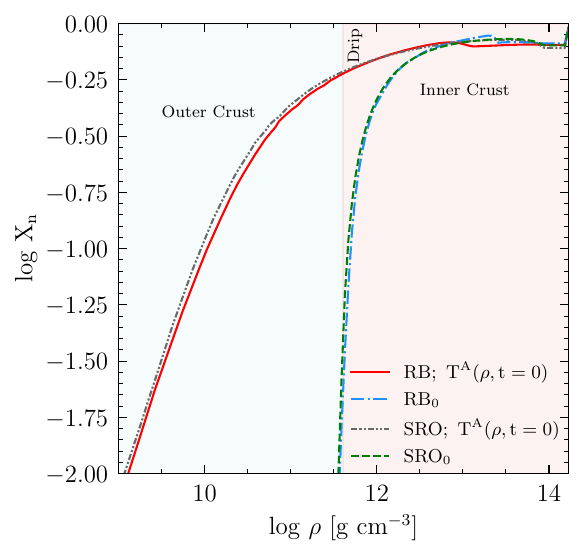}
    \caption{Mass fractions of unbound neutrons in the neo-NS crust as a function of density at $t = 0$.
    For theoretical approaches and local temperature profiles, see Table~\ref{tab:Models}. }
    \label{fig:Compo_Xn}
\end{figure}

\looseness=-1
Figure~\ref{fig:Compo_avAZ} illustrates the composition of the crust at the initial time ($t=0$) in terms of average mass number $\langle A \rangle$ (top panel) and average charge number $\langle Z \rangle$ (bottom panel) as functions of density for the models listed in Table~\ref{tab:Models}.
For RB and SRO, we employ the profile A of the initial temperature; for RB$_0$ and SRO$_0$ this is irrelevant as the composition is fixed to the one at $T_\mathrm{lim}=0.1$~MeV.
In the outermost shells, RB, which uses experimental~\citep{AME2020} and evaluated~\citep{Duflo_PRC_1995} nuclear masses and whose nuclear species distribution is reduced to a single nucleus~\citep{Raduta_AA_2025}, predicts $^{56}$Fe. 
In contrast, SRO, which uses a liquid drop parametrization, favors $^{62}$Fe.
As the density increases, the results of RB$_0$ and SRO$_0$ diverge. 
Particularly for RB and RB$_0$, where the averages are computed over an increasingly wide distribution of species, $\langle A \rangle$ and $\langle Z \rangle$ cease to be representative of any single physical species.
We consider these quantities because they are required by the {\tt NSCool} subroutines. 
The absence of restrictions on nuclear masses implies that, in the SRO$_0$ EoS, deeper shells accommodate more massive nuclei than shallow shells. 
For RB$_0$, closed-shell features are so strong that, over two density domains, $\langle A \rangle$ and $\langle Z \rangle$ decrease with $\rho$. 
Before the crust-core transition, kinetic energy contributions present in RB$_0$ even at temperatures as low as $T=0.1$~MeV are responsible for the replacement of heavy nuclei by the exotic $^{14}$He~\citep{Beznogov_PRC_2026}.
For $\rho \lesssim 10^7$~g/cm$^3$, SRO and RB provide the same compositions as SRO$_0$ and RB$_0$, respectively.
Over $ 10^7 \lesssim \rho~[\mathrm{g/cm^3}] \lesssim 10^{12}$, SRO and RB predict similar values of $\langle A \rangle$ and a non-monotonic density dependence.
The decreasing behavior over $ 10^{7.5} \lesssim \rho~[\mathrm{g/cm^3}] \lesssim 10^{9}$ is due to the onset of other nuclear species: $^4$He and a distribution of light clusters for SRO and RB, respectively. 
Population of these new species also explains why $\langle A \rangle(T) < \langle A \rangle (T=0.1~\mathrm{MeV})$.
As for $T=0.1~\mathrm{MeV}$ and for the same reason, the inner crust of RB is made of light species. 
Finally, nuclear structure effects are washed out at finite temperatures.

Complementary information is provided in Fig.~\ref{fig:Compo_Xn} in terms of the mass fraction of unbound neutrons as a function of density at $t=0$.
Differences in effective interactions and nuclear masses explain why $X_\mathrm{n}$ differs slightly for SRO$_0$ (SRO) compared to RB$_0$ (RB).
$X_\mathrm{n}$ generally increases with temperature. 

Notice that, at finite temperature, unbound neutrons exist at densities orders of magnitude lower than the ``classical'' neutron drip density at $T=0$.
As such, illustration of the drip density in Figs.~\ref{fig:Compo_avAZ}, \ref{fig:Compo_Xn}, \ref{fig:Cv}, and \ref{fig:KT} serves as a reference only.
Furthermore, we refer to unbound neutrons as dripped neutrons without assuming that the density where these neutrons exist corresponds to the density domain of the inner crust at $T=0$.
To emphasize this, we will refer to the unbound neutrons present specifically in the outer crust as ``thermally dripped'' neutrons.

Notice that the thermal drip of protons is also possible, but in our models it was completely negligible. 

\subsection{Specific heat}
\label{ssec:Cv}

\begin{figure}
    \centering
    \includegraphics[width=1.0\linewidth]{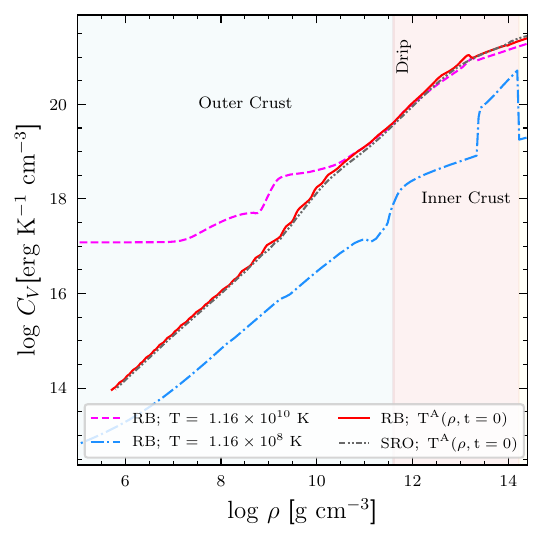}
    \caption{Profiles of specific heat (per unit volume) at constant volume as a function of density for different EoS models and temperature profiles, as indicated in the legend.}
    \label{fig:Cv}
\end{figure}

Equation~\eqref{eq:Ttilde} shows that the thermal evolution of an NS depends on its specific heat at constant volume ($C_\mathrm{V}$).
Accordingly, in Fig.~\ref{fig:Cv}, we investigate the temperature- and density-dependence of this quantity. 
Various EoS models and temperature profiles are considered.
For $T \geq T_\mathrm{lim}$, $C_\mathrm{V}$ is extracted from the \texttt{CompOSE} data by running the \texttt{CompOSE} software, which performs numerical differentiation of the total energy density.
In this way, contributions from nuclei (via nuclear masses, kinetic energy, and excitation spectra), unbound nucleons, electrons, positrons, and photons are implicitly accounted for.
For lower temperatures, we compute $C_\mathrm{V}$ using ``native'' subroutines of \texttt{NSCool}, which include contributions from ions (defined in terms of $\langle A \rangle$ and $\langle Z \rangle$), degenerate neutrons and electrons, under the assumption that the composition is identical to the one at $T_\mathrm{lim}$.
For a fixed temperature, $\log C_\mathrm{V}$ as a function of $\log \rho$ has a complex structure. 
For $T=1.16 \times 10^8$~K, the bump that extends over $10^{13}\,\mathrm{g/cm^3} \lesssim \rho \leq \rho_\mathrm{cc}$ is due to the onset of $^{14}$He. 
For $T=1.16 \times 10^{10}$~K and $\rho \lesssim 10^{7.5}$~g/cm$^3$, $\log C_\mathrm{V}$ is flat. 
This behavior comes from the photon gas and electron-positron pair contributions.
Results of RB and SRO corresponding to the $T^A(\rho)$-profile are similar.
In the first place, this is in line with the state variables' insensitivity with respect to how nuclei are modeled at sub-saturation densities~\citep{Burrows_ApJ_1984}. 
Then, it shows that the uncertainties related to the effective interactions do not manifest in the crust.
To a good approximation, for the $T^A(\rho)$-profile, $\log C_\mathrm{V}$ is a linear function of $\log \rho$.

However, note that the red solid RB curve shows some wobbling, whereas the corresponding gray SRO curve is smooth. 
The other RB curves are smooth as well.
This wobbling in a particular density--temperature domain stems from the way nuclei are treated in the RB EoS compared to the SRO EoS.
Namely, the former is based on an extended nuclear statistical equilibrium (eNSE) model and takes into account shell effects, while the latter is based on a single-nucleus approximation (SNA) plus $^4$He clusters.
The distribution of nuclei in the eNSE model changes appreciably with temperature and density, causing the wobbling in the specific heat. 
The SNA-based composition changes much more smoothly under the same conditions, producing a smoother specific heat.
At higher temperatures (pink dashed curve), the eNSE distributions are wide, the shell effects are washed out, and the specific heat curve is smooth.
As discussed in the previous paragraph, at lower temperatures (blue dash-dotted curve), we do not employ  \texttt{CompOSE} tables; hence, the specific heat curve is also smooth.

\subsection{Thermal conductivity}
\label{ssec:KT}

\begin{figure}
    \centering
    \includegraphics[width=1.0\linewidth]{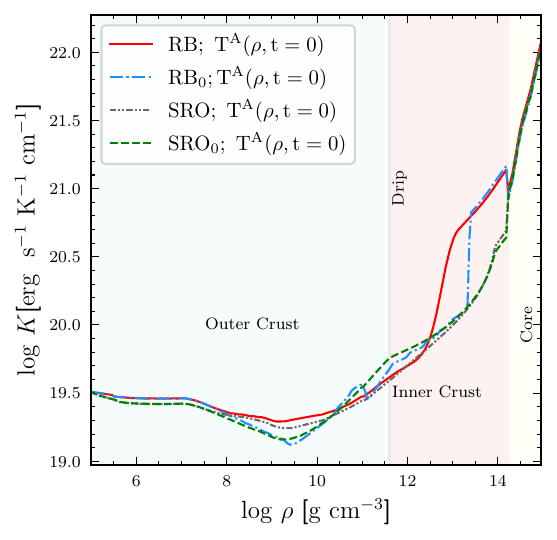}
    \caption{Profiles of thermal conductivity as a function of density at the initial moment of time. }
    \label{fig:KT}
\end{figure}

In Fig.~\ref{fig:KT}, we illustrate the thermal conductivity ($K$) as a function of density as computed by {\tt NSCool} for the models in Table~\ref{tab:Models} at the initial moment of time.
For the details of the calculation, see Appendix~A.2 of Paper~I.
Analyzing the $K(\rho,T)$ dependence allows us to better understand the behavior of the initial temperature profiles in Fig.~\ref{fig:IniProfs}.
More precisely, for a given initial luminosity, models with higher thermal conductivity provide lower temperature gradients, which, based on the initial temperature equality at the bottom of the envelope, get translated into lower temperatures.   

For all models, $\log K$ is a convoluted function of $\log \rho$.
The departure between SRO and SRO$_0$ is smaller than the one between RB and RB$_0$, which is an obvious consequence of employing liquid drop vs experimental masses, respectively.
Because of the onset of $^{14}$He in the innermost shells of the crust, RB and RB$_0$ predict significantly higher thermal conductivities than SRO.

\section{Thermal evolution}
\label{sec:result}

In this section, we present the results of our simulations.
In addition to cooling curves, we follow the evolution of the boundary radius as well as the phase properties of the crust.

\subsection{Cooling Curves} 
\label{sec:cool}

\begin{figure}
  \centering
  \includegraphics[width=\linewidth]{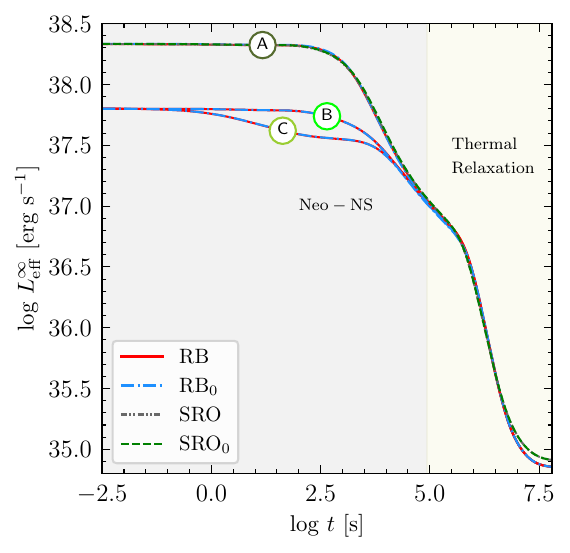}
  \caption{Thermal evolution in terms of redshifted effective luminosity as a function of time for $t < 1$~yr.
  Predictions for various models of the crust composition and profiles of initial luminosity.
  For details, see Table~\ref{tab:Models} and Fig.~\ref{fig:IniProfs}.
  Different cooling stages are indicated in the background.}
  \label{fig:Leff_vs_t_short}
\end{figure}

\begin{figure}
  \centering
  \includegraphics[width=\linewidth]{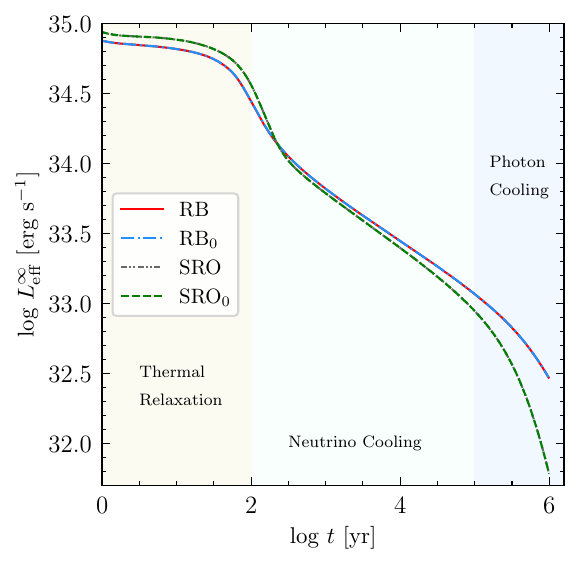}
  \caption{Thermal evolution in terms of redshifted effective luminosity as a function of time for $0 \leq \log t\,[\rm{yr}] \leq 6$.
  Predictions for various models of the crust composition and the $L_0^{A}(\rho)$ profile of initial luminosity.
  For details, see Table~\ref{tab:Models} and Fig.~\ref{fig:IniProfs}.
  Different cooling stages are indicated in the background.}
  \label{fig:Leff_vs_t_long}
\end{figure}

\begin{figure}
  \centering
  \includegraphics[width=\linewidth]{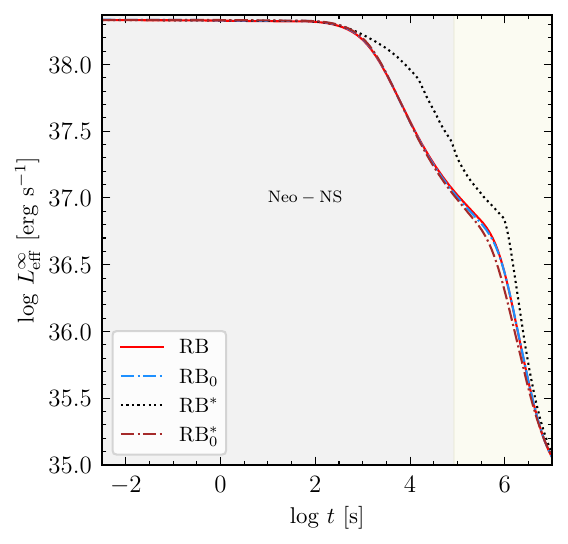}
  \caption{The same as in Fig.~\ref{fig:Leff_vs_t_short}.
  	Calculations for RB and RB$_0$ models and various recipes for calculating $C_\mathrm{V}$.
  	See text for details. }
  \label{fig:Leff_details}
\end{figure}

\looseness=-1
Figure~\ref{fig:Leff_vs_t_short} demonstrates the short-term thermal evolution of $1.4~M_{\odot}$ NSs whose crusts are modeled by the RB and SRO EoSs.
As indicated on the figure, $t < 1$~yr encompasses the entire neo-NS stage along with the early thermal relaxation stage.
Results for the temperature-dependent compositions (RB and SRO) are compared with those obtained using fixed compositions (RB$_0$ and SRO$_0$) for each of the three initial luminosity profiles shown in Fig.~\ref{fig:IniProfs}. 
Over $0.3 \lesssim t~[s] \lesssim 10^2$, $L_{\mathrm{eff}}^{\infty;B}$ stays flat while $L_{\mathrm{eff}}^{\infty;C}$ steadily decreases.
This behavior replicates the behaviors of $L_0$ over $\rho_\mathrm{b} \leq \rho \lesssim 10^7$~g/cm$^{3}$.
The plateau of $L_{\mathrm{eff}}^{\infty;A}$ lasts almost as long as the plateau of $L_{\mathrm{eff}}^{\infty;B}$.
This shows that the higher the initial luminosity, the sooner $L_{\mathrm{eff}}^{\infty}$ drops.
At $t \approx 10^{4}$~s, the small discrepancies between the two low-luminosity initial profiles are washed out.
At $t \approx 3\times 10^{4}$~s, all differences among the initial luminosity profiles are washed out. 
Indeed, from this moment on, all $L_{\mathrm{eff}}^{\infty}(t)$ curves are identical. 
We note that in Paper~I, which only allowed for the temperature dependence
of the mechanical structure of the layers with $10^5 \leq \rho~[\mathrm{g/cm^3}] \leq 10^{11}$, the memory of the initial luminosity profiles was washed out slightly earlier, i.e., at $\approx 10^4$~s.

The curves corresponding to RB$_0$ and SRO$_0$, which do not allow the composition to change with temperature, are indistinguishable from the curves corresponding to RB and SRO until $t \sim 10^7$ s.
Moreover, no differences are obtained between $L_{\mathrm{eff}}^{\infty}$ predicted by RB and SRO.
The obvious conclusion is that the crust composition has a negligible effect on the thermal evolution during the neo-NS phase. 
This should not come as a surprise since these models feature minor differences in thermal conductivities and have almost identical specific heat capacities, with the exception of the innermost shells.
Due to the thermal diffusion timescales from the inner crust to the surface, those differences in the inner crust cannot be observed on the surface during the neo-NS phase.

Figure~\ref {fig:Leff_vs_t_long} investigates the long-term thermal evolution of the same NSs considered in Fig.~\ref{fig:Leff_vs_t_short}.
As before, the predictions of RB$_0$ (SRO$_0$) are identical to those of RB (SRO).
However, the results of RB are no longer identical with those of SRO. 
Indeed, during the thermal relaxation and neutrino cooling epochs, RB and SRO are slightly different, while starting from the photon cooling epoch, RB curves stay much warmer than SRO curves.
The differences during the thermal relaxation phase stem from two sources.
First, the luminosity throughout the plateau phase ($1 \lesssim t~[\mathrm{yr}] \lesssim 100$) is partially determined by the suppression of plasmon decay neutrinos (see Sec.~5.2 and Appendix~A.4 of Paper~I).
The plasma frequency for RB and RB$_0$ in the inner crust is slightly lower than that for SRO and SRO$_0$.
Consequently, plasmon decay neutrinos turn off slightly later for RB and RB$_0$, thereby making the surface luminosity for these models slightly lower. 
Second, the timing of the crust-core thermalization (signifying the transition to the neutrino cooling phase at $t \sim 100$~yr) is determined by the timescale of heat diffusion through the crust.
As demonstrated in Fig.~\ref{fig:KT}, the thermal conductivity of RB and RB$_0$ in the inner crust significantly exceeds that of SRO and SRO$_0$.
As a result, crust-core thermalization for RB and RB$_0$ occurs earlier.
The difference observed in the photon cooling epoch comes from the very high specific heat capacity of $^{14}$He that is present in the innermost shells of the crusts of RB models, see Fig.~\ref{fig:Cv}.

In Fig.~\ref{fig:Leff_details} we offer additional information about the neo-NS phase.
The cooling curves RB$^*$ and RB$^*_0$ are counterparts of the cooling curves RB and RB$_0$ where, instead of using the accurate $C_\mathrm{V}$ computed by numerical differentiation of the energy density, we compute $C_\mathrm{V}$ via the subroutines of \texttt{NSCool}.
Those subroutines calculate the contribution of dripped neutrons to the specific heat capacity ($C_\mathrm{V;n}$) within the degenerate Fermi gas limit and ignore contributions from thermally excited states of the nuclei.
For zero temperature EoSs, which are employed in most NSs' cooling studies, the degenerate gas approximation is accurate everywhere except the outermost shells of the inner crust, where the density of dripped neutrons is low; discarding internal excitations is fully justified. 
Finite temperatures increase the number of dripped neutrons in the density domain that, at $T=0$, corresponds to the inner crust and lead to thermal dripping at densities lower than the drip density, see Fig.~\ref{fig:Compo_Xn}. 
Finite temperatures also allow for the population of nuclear excited states, whose numbers increase with the excitation energy and nuclear mass.

The use of the degenerate approximation for neutrons in the temperature-density domain where they are not degenerate obviously leads to a \emph{significant} overestimation of the neutron heat capacity (1-2 orders of magnitude, \citealt{Constantinou_PRC_2014}).
Discarding excited states leads to an underestimation of the total heat capacity.
Now, for $t \lesssim 1$~yr the overestimated heat capacity of neutrons in the inner crust does not produce any effect on the surface, see the discussion below regarding RB$_0^*$ curve. 
At later times, the inner crust is colder and the degenerate gas limit is applicable.
Thus, the results of Paper~I remain valid.

The situation changes noticeably when the thermally dripped neutrons in the outer crust are taken into account.
If their heat capacity is overestimated, this is observable on the surface, as demonstrated by the RB$^*$ curve.
RB$^*$ shows a substantially slower cooling than RB and RB$_0$, which means that the overestimation of $C_\mathrm{V;n}$ in the outer crust has a much stronger impact than neglecting the contributions of the excited states. 
RB$^*_0$ corresponds to the RB composition at $T=T_\mathrm{lim}$ and $C_\mathrm{V;n}$ computed in the degenerate gas approximation; it is a proxy for the procedure adopted in Paper~I (modulo the contribution of the nuclear excited states, which were actually taken into account in Paper~I).
It does not have thermally dripped neutrons in the outer crust, nor does it account for their heat capacity. 
It does overestimate the heat capacity of the neutrons in the inner crust.
Yet, we see that it cools much faster than RB$^*$ and almost the same as RB and RB$_0$.
This proves that the overestimation of the heat capacity of the inner crust is not observable, while the overestimation of the heat capacity of the outer crust is quite noticeable.
The difference comes from the heat diffusion timescale from the given point inside the star to the surface (see Paper~I). 
For the outer crust $t_\mathrm{diff} \lesssim t = 1$~yr, for the inner crust $t_\mathrm{diff} \gtrsim t=1$~yr.

The fact that, over the interval $10^{3.5} \lesssim t~[s] \lesssim 10^6$, RB$^*_0$ predicts slightly faster cooling than RB/RB$_0$ demonstrates the impact of the heat capacity of thermally dripped neutrons in the outer crust and the contribution of the nuclear excited states to $C_\mathrm{V}$.
As one can see, these contributions combined have little influence on the cooling curves.

\subsection{Contraction of cooling down NSs}
\label{sec:evolve}

\begin{figure}
    \centering
    \includegraphics[width=1.0\linewidth]{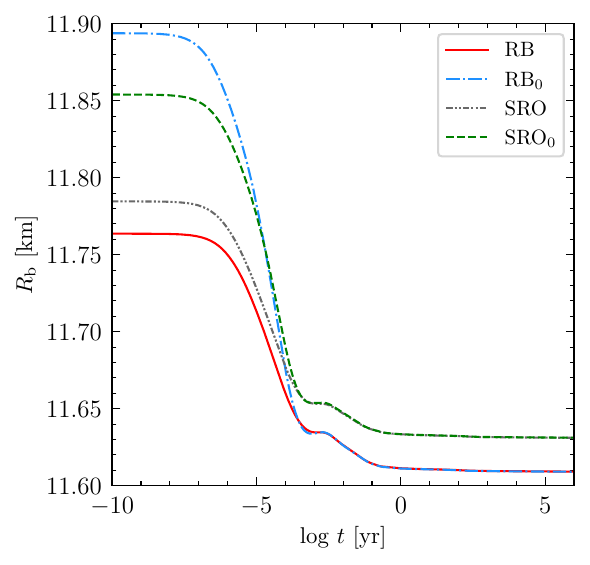}
    \caption{Time evolution of the boundary between the NS outer crust and the bottom of the envelope.
    Results corresponding to fixed and temperature-dependent crust compositions are confronted for two EoSs.
    All curves correspond to initial luminosity profile ``A''.}
    \label{fig:Rb_vs_t}
\end{figure}

In Fig.~\ref{fig:Rb_vs_t} we show the time evolution of the boundary between the NS outer crust and the bottom of the envelope, whose density at $t=0$ is $\rho_\mathrm{b}=10^5$~g/cm$^3$. 
As expected, NSs shrink as they cool.
For some time after the beginning of the simulation, the $R_\mathrm{b}$ values are constant and their ordering replicates the ordering of $T_0$ in the inner crust, see the bottom panel in Fig.~\ref{fig:IniProfs}.
This suggests that the stellar radius is dictated by the thickness of the inner crust, which increases with temperature.
Confrontation with the cooling curves in Fig.~\ref{fig:Leff_vs_t_short} shows that the boundary radii stay constant roughly as long as the corresponding surface luminosities ($L_{\mathrm{eff}}^{\infty}$) stay constant, i.e., for $\approx 300$~s.

The constancy of $R_\mathrm{b}$ persists longer for RB than for RB$_0$. 
Also, $\left\{R_\mathrm{b}(t=0; \mathrm{RB_0})-R_\mathrm{b}(t=0; \mathrm{RB})\right\}$ exceeds $\left\{R_\mathrm{b}(t=0; \mathrm{SRO_0})-R_\mathrm{b}(t=0; \mathrm{SRO})\right\}$.
This can be understood considering that the difference between the thermal conductivities of SRO and SRO$_0$ is less than the corresponding difference between RB and RB$_0$, see Fig.~\ref{fig:KT}. 
The small plateaus around 8~hr come from the fact that both RB and SRO EoSs have temperature-density domains where $\left(\partial P/\partial T \right)_n<0$.
While this is insufficient to actually expand the star, it nevertheless pauses contraction for some time.
From 1~yr onwards, $R_\mathrm{b}$ values remain constant, with higher values for SRO than for RB, reflecting the stiffnesses of the two EoSs.

\subsection{Crust crystallization}
\label{sec:Crystal}

\begin{figure*}
    \centering
    \includegraphics[width=1.0\linewidth]{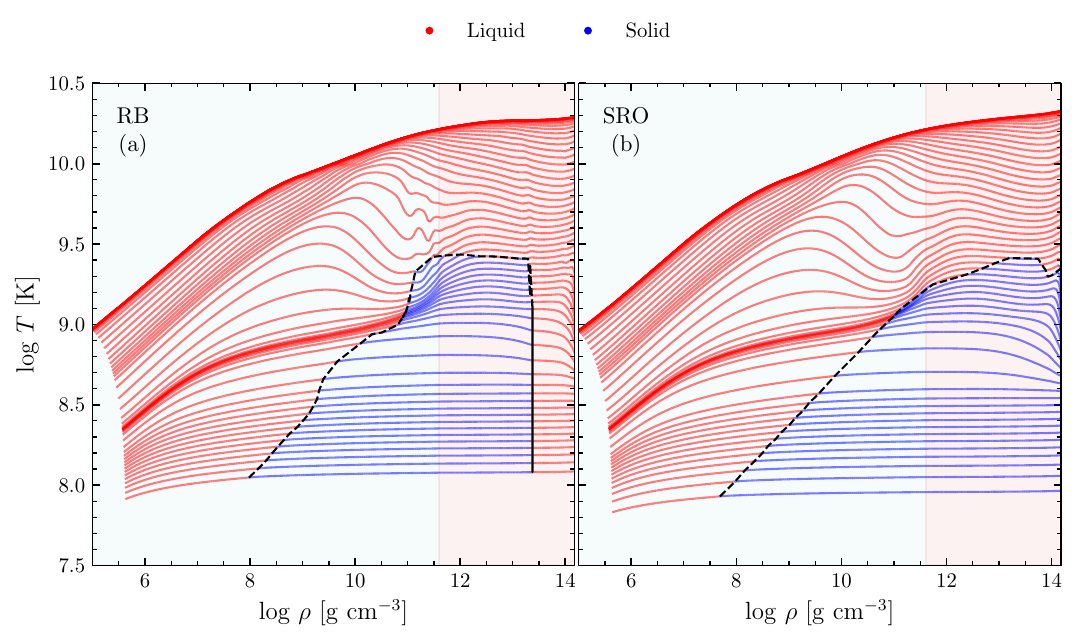}
    \caption{Local temperature profiles as a function of density at various instances in time during the entire cooling process.
    In both panels, the curves are plotted at identical times. 
    The hottest (topmost) curves correspond to $\log_{10} t [\mathrm{s}] = -2.5$, and
    the coldest  (lowest) curves correspond to $\log_{10} t [\mathrm{s}] = 13$;
    the remaining curves are sampled uniformly in $\log_{10} t$ within these boundaries. 
    In total, there are 63 curves in each panel, which corresponds to a logarithmic step of 0.25.
    Different panels correspond to different crust compositions; for legend, see Table~\ref{tab:Models}.
    All panels correspond to the initial luminosity profile ``A''.
    $T-\rho$ domains where matter is solid (liquid) are marked in blue (red).}
    \label{fig:Crystal}
\end{figure*}

Figure~\ref{fig:Crystal} illustrates local temperature profiles as a function of density at various instances in time during the entire cooling process, as described in the caption.
Results corresponding to RB$_0$ are not shown as they are identical to those corresponding to RB.
If not explicitly mentioned otherwise, comments in this paragraph refer to both the RB and SRO EoSs.

One can see that for $t \lesssim 10~\mathrm{s}$, the temperature profiles show a monotonic decrease from the crust-core interface toward the outer boundary, while more complex $T(\rho)$ dependencies are obtained at later times.
Over $15 \lesssim t~[\mathrm{s}] \lesssim 2 \times 10^4$, $T(\rho)$ exhibits one maximum that is localized either in the outer shells of the inner crust or the inner shells of the outer crust and whose position shifts toward lower densities as the time proceeds.
Over $0.001 \lesssim t~[\mathrm{yr}] \lesssim 0.18$, the profiles of local temperature show two maxima, one situated in the inner crust and the other one in the outer crust; the position of the latter shifts toward lower densities.
For the RB EoS, $T(\rho)$ also shows some small bumps.
Most likely, they are the outcome of irregularities in the $C_\mathrm{V}(\rho,T)$ extracted from the \texttt{CompOSE} tables, see Fig.~\ref{fig:Cv} and the discussion in Sec.~\ref{ssec:Cv}.
Notably, these bumps are not present for the SRO profiles and the SRO $C_\mathrm{V}(\rho,T)$ curve in Fig.~\ref{fig:Cv} is smooth.
For $0.3 \lesssim t~[\mathrm{yr}] \lesssim 6$, the local temperature in the inner crust is constant, while in the outer crust it decreases.
For $10 \lesssim t~[\mathrm{yr}] \lesssim 180$, $T(\rho)$ manifests a wide local maximum situated around the neutron drip line and the temperature at the crust core boundary is lower than the temperature in the inner crust.
This situation is due to the propagation of the cooling wave from the core, which cools faster than the crust because of its high neutrino luminosity.

Similar profiles were obtained in classical papers on NS cooling~\citep{Gnedin_MNRAS_2001,Page_WS_2013} as well as in Paper~I.
From $t \approx 300\,\mathrm{yr}$ onward, the profiles are almost flat, signifying that the star has become almost isothermal inside.
Note that formally the redshifted temperature is constant when the star is isothermal, but the variation of the redshift across the entire crust is relatively small.
The visual ``clustering'' of the curves is partially due to plateaus in the cooling curves and partially due to the way in which we sample the ages for plotting.

As the temperature decreases, we expect the inhomogeneous matter in the crust to switch from a liquid phase to a solid one.
Following \cite{Potekhin_2010}, we employ two criteria for distinguishing between the liquid and solid phases.
The first one is the classical one-component plasma melting criterion,
\begin{equation}
    k_B T_{\mathrm m} (n_{\mathrm i})=\frac{\left(\langle Z \rangle e\right)^2}{a_{\mathrm i} \Gamma_{\mathrm m}},
    \label{eq:Tm}
\end{equation}
where $n_{\mathrm i}$ is the number density of ions, $a_{\mathrm i}=\left( 4 \pi n_{\mathrm i}/3\right)^{-1/3}$ is the ion-sphere radius and $\Gamma_{\mathrm m}=175$ is the Coulomb coupling strength at the melting point.
Eq.~\eqref{eq:Tm} defines the melting temperature as a function of density.

The second criterion regards ``quantum melting''. 
It is defined by the value of the ion density parameter,
\begin{equation}
    R_S=a_{\mathrm i} m_{\mathrm i} \frac{\left(\langle Z \rangle e\right)^2}{\hbar^2},
    \label{eq:Rs}
\end{equation}
where $m_{\mathrm i}$ is the ion mass. 
$R_S$ depends on the ion density and the temperature enters only via $\langle Z \rangle$. 
The critical value of $R_S$ is taken as 140, which defines the corresponding critical density $\rho_{\mathrm m}$. 
If the ion density exceeds this critical density, the ion liquid does not crystallize no matter the temperature.
In practice, only $^{14}$He layers present in the RB EoS can ``quantum melt''.

If $T < T_{\mathrm m}(\rho)$ and $\rho < \rho_{\mathrm m}$, we consider that the matter is solid.
Corresponding segments of curves are marked in blue.
Otherwise, the matter is in the liquid phase and corresponding segments are marked in red.

Figure~\ref{fig:Crystal} shows that for RB (SRO) EoS and $t \lesssim 0.01~\mathrm{yr}$ ($t \lesssim 0.06~\mathrm{yr}$) the whole crust is liquid.
The first solid layers appear at $T \approx 10^{9.4}~\mathrm{K} \approx 0.2$~MeV, which occurs at $t \approx 0.018~\mathrm{yr}$ ($t \approx 0.1~\mathrm{yr}$) for RB (SRO).
The uncertainty in the crystallization onset age comes from the temporal discretization of the curves in the figure. 
For RB (SRO), these first solid layers are located in the deep shells of the outer crust (intermediate shells of the inner crust) and correspond to the density domain $10^{11} \lesssim \rho~[\mathrm{g/cm^3}] \lesssim 10^{11.4}~$ ($10^{13} \lesssim \rho~[\mathrm{g/cm^3}] \lesssim 10^{13.7}$).
For $T \approx 10^{9.4}~\mathrm{K}$, the solid phase occupies a significant (narrow) region of the inner crust for RB (SRO).
The higher values of $T_{\mathrm{m}}$ around $\rho_{\mathrm{n;\,drip}}$ in case of RB are due to the more massive nuclei, see Fig.~\ref{fig:Compo_avAZ}.
According to RB, the layers with $10^{13.2}~\mathrm{g/cm^3}\leq \rho \leq \rho_{\mathrm{cc}}$, which correspond to the innermost shells of the inner crust, remain liquid up to $t=3 \times 10^5~\mathrm{yr}$.
In contrast, SRO predicts that the liquid layers in the innermost crust persist only up to $t \lesssim 0.3~\mathrm{yr}$ and correspond to $10^{13.7}~\mathrm{g/cm^3} \lesssim \rho \leq \rho_{\mathrm{cc}}$.
Again, the striking difference between the RB and SRO EoSs comes from the differences in the composition.
In case of the RB EoS, the deepest layers of the crust are made of $^{14}$He, while in case of the SRO EoS, they are made of heavy nuclei, see Fig.~\ref{fig:Compo_avAZ}.
An order of magnitude estimation based on Eq.~\eqref{eq:Tm} shows that $T_{\mathrm m}^{\mathrm {RB}}/T_{\mathrm m}^{\mathrm {SRO}} \approx 1/100$.

As the star cools down, the solid phase gradually extends in the outer crust.
Still, for the time duration considered in Fig.~\ref{fig:Crystal}, i.e., $t \leq 3 \times 10^5~\mathrm{yr}$, the shells with $\rho \lesssim 10^8~\mathrm{g/cm^3}$ remain liquid.
The outcomes of RB and SRO also differ at the latest  times considered.
According to RB, layers with $\rho \lesssim 10^{8}~\mathrm{g/cm^3}$ are liquid, whereas for SRO, the liquid phase corresponds to $\rho \lesssim 10^{7.5}~\mathrm{g/cm^3}$.
This is because the SRO EoS results in faster cooling during the photon phase (see Fig.~\ref{fig:Leff_vs_t_long}).
Indeed, at $t \sim 3 \times 10^5~\mathrm{yr}$, the local temperature of the solid layer is $10^{8.1}~\mathrm{K}$ ($10^{7.95}~\mathrm{K}$) for RB (SRO).

\subsection{On the effects of $^1\mathrm{S}_0$ neutron pairing}
\label{ssec:sf}

\begin{figure}
    \centering
    \includegraphics[width=1.0\linewidth]{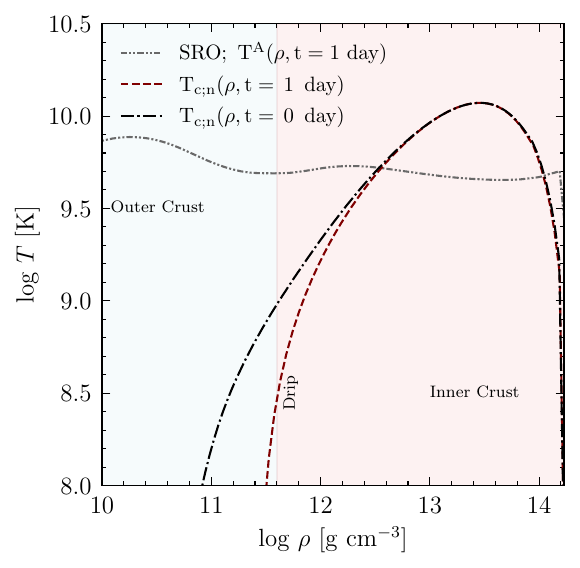}
    \caption{The profile of local temperature and the profiles of critical temperature for $^1\mathrm{S}_0$ neutron pairing~\citep{Ding_2016} inside the crust.}
    \label{fig:SF}
\end{figure}

The impact of $^1\mathrm{S}_0$ neutron pairing on the long-term cooling of NSs is a well-studied question (see, e.g., \citealt{1994ApJ...425..802L,Gnedin_MNRAS_2001,Yakovlev_2001,Page_ApJ_2004,Page_ApJ_2009,Fortin_PRC_2010,2020ApJ...898..125P}).
Here, we briefly discuss how the pairing affects the cooling during the neo-NS phase.
In short, the effect is negligible.
From Fig.~\ref{fig:SF} one can see, first, that at the age of 1 day the temperature profile intersects with the neutron $^1\mathrm{S}_0$ critical temperature ($T_\mathrm{c;n}$; \citealt{Ding_2016}) only in the inner crust. 
Thus, as discussed in relation to Fig.~\ref{fig:Leff_details}, any changes in the temperature profile inside the star induced by superfluidity will be unobservable on the surface during the neo-NS phase.
Second, while the $T_\mathrm{c;n}$ profile does extend into the outer crust, this happens only for the age $t \lesssim 1$~day when the temperature in the outer crust is much higher than $T_\mathrm{c;n}$ in the same region. 
Consequently, thermally dripped neutrons in the outer crust remain unpaired as long as they exist, which means, again, that the cooling curves are not affected during the neo-NS phase.

This conclusion is straightforward, as the typical timescale for crustal superfluidity to impact the cooling curves is a few decades, whereas here we consider the evolution at earlier times ($t < 1$~yr). 

Nevertheless, it would be useful to perform explicit calculations to determine how superfluidity affects the temperature profiles inside the inner crust.
However, this is problematic. 
\cite{Raduta_AA_2025} noted that inhomogeneous warm matter (finite-temperature crustal matter) cannot be ``split'' into thermodynamically-consistent (sub-)components. 
This implies that it is not possible to self-consistently ``extract'' dripped neutrons, modify their thermodynamics/kinetics according to their pairing, and then reintroduce them into the system.
The only way to treat them correctly would be to include pairing directly into the computation of the EoS.
A few such treatments exist, e.g., \cite{Fortin_PRC_2010,Burrello_PRC_2015}, but this is a complicated task that is far outside the scope of the current paper.

\section{Summary and conclusions} 
\label{sec:Concl}

This work builds upon that of \citet{Beznogov_ApJ_2020}, who investigated the thermal evolution of young isolated neutron stars during the first $\sim 3$~yr after birth. 
In addition to accounting for temperature effects on the $P(\rho)$ relation in the envelope, we accounted for temperature-induced modifications of the entire crust  both in terms of the $P(\rho,T)$ dependence and the composition.  
Our version of {\tt NSCool} uses temperature-dependent compositions and $P(\rho)$-relations for the entire crust, adjusting the mechanical structure at each timestep during the evolution.
The EoS information is extracted from general-purpose \texttt{CompOSE} tables, assuming that the matter is in beta-equilibrium and that this corresponds to the same relation among neutron, proton and electron chemical potentials as at $T=0$, but with the values corresponding to the current local temperature profile.

\looseness=-1
We conducted simulations for three profiles of initial local luminosity and two families of EoS models. Our results show that the memory of the initial profiles is washed out in $\sim 3 \times 10^4$~s, which is slightly longer than the time it took for neo-NSs in Paper~I to forget the initial conditions. 
Comparing cooling curves obtained using the same $P(\rho,T)$ relation and specific heat capacity but different compositions reveals that, while the composition changes noticeably, the thermal conductivity in the outer crust is almost the same and, thus, there are no differences in cooling.
Conversely, when self-consistent composition and $P(\rho,T)$ are implemented into simulations, 
temperature-induced modifications of the composition manifest through the total heat capacity.
In particular, by slightly increasing the heat capacity of the outer crust compared to its value in cold NSs, thermally dripped neutrons are responsible for a bit slower cooling during the thermal relaxation phase. 
Additional composition effects manifest whenever the EoS model favors nucleation of light species in the deepest layers of the warm crust, as demonstrated by \cite{Beznogov_PRC_2026}. 
In these situations, the value of $L_\mathrm{eff}^{\infty}$ in the thermal relaxation phase is reduced and the cooling in both neutrino and, especially, photon cooling phases is slowed down. 
The latter effect is the obvious consequence of the increased heat capacity. 
During the neo-NS phase, $L_\mathrm{eff}^{\infty}$ is sensitive to the properties of the outer crust, making it insensitive to $^1S_0$ neutron pairing.

Allowing for mechanical structure adjustments in the cooling process revealed that during the first $\sim 1$~yr the radius of the star decreased by 0.25~km, with the most important modification occurring for $10^{-7} \lesssim t\,[\mathrm{yr}] \lesssim 10^{-1}$. 
We also tracked the crystallization of the crust and investigated its EoS dependence.
For EoS models favoring massive nuclei, crust crystallization begins at $t \sim 0.1~\mathrm{yr}$, when the local temperature is approximately 0.2~MeV, and proceeds outward from the deepest layers of the inner crust.
In contrast, in models that allow for light species in the deepest layers, these layers remain liquid for an extended period of time.
According to our melting criteria, layers with $\rho \lesssim 10^{7.5} \, \mathrm{g/cm^3}$ remain liquid even at $t \sim 3.2 \times 10^5 \,\mathrm{yr}$.

Our models can maintain near-Eddington luminosity for approximately the same duration (several hundred seconds) as the baseline model of Paper~I.
The entire neo-NS phase lasts approximately a day.
Thus, the chances of detecting an NS during this phase are small. 
Indeed, after a CCSN, the ejected material has to become transparent to the thermal emission of a neo-NS, which takes, in the most optimistic case, of the order of a year (Sec.~6.1 of Paper~I).
Obviously, this makes the detection of a neo-NS after a CCSN problematic.
The youngest known (and not yet finally confirmed) NS is the one in SN1987A.
It is almost 40 years old and was tentatively detected indirectly by the radio emission from the heated dust \citep{Cigan,2020ApJ...898..125P}
and not by its thermal X-ray emission from the surface\,\footnote{Now it is also seen in the infrared by the James Webb Space Telescope \citep{Larsson_2025}; still, no direct detection in X-rays.}.

The chances of detecting a neo-NS after an accretion-induced collapse of a WD are better, as there are significantly less ejecta.
However, they can be confused with other thermal transients having similar timescales.
Finally, the best chance to explicitly detect a neo-NS is after a binary NS merger seen off-axis.
In that case, there are also much less ejecta than after a CCSN, and the material should become transparent on a timescale of days.
For more details and references, see Sec.~6 of Paper~I.

We have ignored rotation and magnetic fields in our calculations.
Both aspects were discussed in Paper~I, Secs.~5.4 and 5.6, respectively.
We do not expect any differences here due to the temperature-dependent inner crusts.

\vspace{0.3cm}

Employed software: {\ttfamily neo-NSCool} with temperature-dependent composition and EoS 
(not publicly available, built upon \citealt{Page_NSCool_1d} and \citealt{Beznogov_ApJ_2020}),
{\ttfamily python} \citep{python},
{\ttfamily matplotlib} \citep{Hunter:2007},
{\ttfamily f90wrap} \citep{f90wrap},
{\ttfamily numpy} \citep{harris2020array},
{\ttfamily scipy} \citep{scipy}.

Data availability: the data underlying Figs.~\ref{fig:Leff_vs_t_short}, \ref{fig:Leff_vs_t_long}, and \ref{fig:Leff_details} is provided in \cite{Neo-NSs_data} [DOI: \href{https://doi.org/10.5281/zenodo.23343164}{10.5281/zenodo.23343164}].

\begin{acknowledgements}
S.C, M.B. and A.R. acknowledge support from a grant from the Ministry of Education and Research, CNCS/CCCDI–UEFISCDI, Project No. PN-IV-P1-PCE-2023-0324 and partial support from Project No. PN 23 21 01 02.
D.P.’s work is supported by a UNAM-DGAPA grant PAPIIT-IN114424.
\end{acknowledgements}

\bibliography{NeoNS_v3}
\bibliographystyle{aa_link}
\end{document}